\documentclass[pdflatex,sn-mathphys]{sn-jnl}

\jyear{2023}%

\theoremstyle{thmstyleone}%
\theoremstyle{thmstyletwo}%

\theoremstyle{thmstylethree}%
\usepackage[numbers]{natbib}
\usepackage{amsmath,amssymb}
\usepackage{graphicx}
\usepackage{caption}
\usepackage{url}
\usepackage{setspace}
\usepackage{multibib}
\newcites{supp}{Methods References}

\usepackage{graphicx}	
\usepackage{amsmath}	

\usepackage{amssymb}	
\usepackage[T1]{fontenc} 
\usepackage{academicons}
\usepackage{xcolor}
\usepackage{multirow}
\usepackage{float}
\usepackage{subcaption}
\usepackage[superscript,biblabel]{cite}
\usepackage{rotating}

\usepackage{newtxtext,newtxmath}

\begin{document}

\title[Dwarf galaxies reionized the Universe]{Most of the photons that reionized the Universe came from dwarf galaxies}

\author*[1]{\fnm{Hakim} \sur{Atek}}
\author[2]{\fnm{Ivo} \sur{Labb\'e}}
\author[3]{\fnm{Lukas J.} \sur{Furtak}}
\author[1]{\fnm{Iryna} \sur{Chemerynska}}
\author[4]{\fnm{Seiji} \sur{Fujimoto}}
\author[5]{\fnm{David J.} \sur{Setton}}
\author[6]{\fnm{Tim B.} \sur{Miller}}
\author[7,8]{\fnm{Pascal} \sur{Oesch}}
\author[5]{\fnm{Rachel} \sur{Bezanson}}
\author[5]{\fnm{Sedona H.} \sur{Price}}
\author[9]{\fnm{Pratika} \sur{Dayal}}
\author[2]{\fnm{Adi} \sur{Zitrin}}
\author[9]{\fnm{Vasily} \sur{Kokorev}}
\author[10]{\fnm{John R.} \sur{Weaver}}
\author[8]{\fnm{Gabriel} \sur{Brammer}}
\author[11]{\fnm{Pieter} \sur{van Dokkum}}
\author[12,13]{\fnm{Christina C.} \sur{Williams}}
\author[10]{\fnm{Sam E.} \sur{Cutler}}
\author[15]{\fnm{Robert} \sur{Feldmann}}
\author[16,17]{\fnm{Yoshinobu} \sur{Fudamoto}}
\author[14]{\fnm{Jenny E.} \sur{Greene}}
\author[18,19,20]{\fnm{Joel} \sur{Leja}}
\author[21]{\fnm{Michael V.} \sur{Maseda}}
\author[22]{\fnm{Adam} \sur{Muzzin}}
\author[23]{\fnm{Richard} \sur{Pan}}
\author[24,25]{\fnm{Casey} \sur{Papovich}}
\author[26]{\fnm{Erica J.} \sur{Nelson}}
\author[2]{\fnm{Themiya} \sur{Nanayakkara}}
\author[27]{\fnm{Daniel P.} \sur{Stark}}
\author[28]{\fnm{Mauro} \sur{Stefanon}}
\author[29,30]{\fnm{Katherine A.} \sur{Suess}}
\author[18,19,20]{\fnm{Bingjie} \sur{Wang}}
\author[8,10]{\fnm{Katherine E.} \sur{Whitaker}}

\affil[1]{Institut d’Astrophysique de Paris, CNRS, Sorbonne Universit\'{e}, 98bis Boulevard Arago, 75014, Paris, France}
\affil[2]{Centre for Astrophysics and Supercomputing, Swinburne University of Technology, Melbourne, VIC 3122, Australia}
\affil[3]{Physics Department, Ben-Gurion University of the Negev, P.O. Box 653, Be'er-Sheva 84105, Israel}
\affil[4]{Department of Astronomy, The University of Texas at Austin, Austin, TX 78712, USA}
\affil[5]{Department of Physics and Astronomy and PITT PACC, University of Pittsburgh, Pittsburgh, PA 15260, USA}
\affil[6]{Center for Interdisciplinary Exploration and Research in Astrophysics (CIERA) and Department of Physics \& Astronomy, Northwestern University, IL 60201, USA}
\affil[7]{Department of Astronomy, University of Geneva, Chemin Pegasi 51, 1290 Versoix, Switzerland}
\affil[8]{Cosmic Dawn Center (DAWN), Niels Bohr Institute, University of Copenhagen, Jagtvej 128, K{\o}benhavn N, DK-2200, Denmark}
\affil[9]{Kapteyn Astronomical Institute, University of Groningen, P.O. Box 800, 9700 AV Groningen, The Netherlands}
\affil[10]{Department of Astronomy, University of Massachusetts, Amherst, MA 01003, USA}
\affil[11]{Department of Astronomy, Yale University, New Haven, CT 06511, USA}
\affil[12]{NSF’s National Optical-Infrared Astronomy Research Laboratory, 950 N. Cherry Avenue, Tucson, AZ 85719, USA}
\affil[13]{Steward Observatory, University of Arizona, Tucson, AZ 85721, USA}
\affil[14]{Department of Astrophysical Sciences, Princeton University, 4 Ivy Lane, Princeton, NJ 08544}
\affil[15]{Institute for Computational Science, University of Zurich, CH-8057 Zurich, Switzerland}
\affil[16]{Waseda Research Institute for Science and Engineering, Faculty of Science and Engineering, Waseda University, 3-4-1 Okubo, Shinjuku, Tokyo 169-8555, Japan}
\affil[17]{National Astronomical Observatory of Japan, 2-21-1, Osawa, Mitaka, Tokyo, Japan}
\affil[18]{ Department of Astronomy \& Astrophysics, The Pennsylvania State University, University Park, PA 16802, USA}
\affil[19]{Institute for Computational \& Data Sciences, The Pennsylvania State University, University Park, PA 16802, USA}
\affil[20]{Institute for Gravitation and the Cosmos, The Pennsylvania State University, University Park, PA 16802, USA}
\affil[21]{Department of Astronomy, University of Wisconsin, 475 N. Charter Street, Madison, WI 53706, USA}
\affil[22]{Department of Physics and Astronomy, York University, 4700 Keele Street, Toronto, ON, M3J 1P3, Canada}
\affil[23]{Department of Physics and Astronomy, Tufts University, 574 Boston Ave., Medford, MA 02155, USA}
\affil[24]{Department of Physics and Astronomy, Texas A\&M University, College Station, TX, 77843-4242 USA}
\affil[25]{George P.\ and Cynthia Woods Mitchell Institute for Fundamental Physics and Astronomy, Texas A\&M University, College Station, TX, 77843-4242 USA}
\affil[26]{Department for Astrophysical and Planetary Science, University of Colorado, Boulder, CO 80309, USA}
\affil[27]{Steward Observatory, University of Arizona, 933 N Cherry Ave, Tucson, AZ 85721 USA}
\affil[28]{Departament d'Astronomia i Astrof\`isica, Universitat de Val\`encia, C. Dr. Moliner 50, E-46100 Burjassot, Val\`encia,  Spain}
\affil[29]{Department of Astronomy and Astrophysics, University of California, Santa Cruz, 1156 High Street, Santa Cruz, CA 95064 USA}
\affil[30]{Kavli Institute for Particle Astrophysics and Cosmology and Department of Physics, Stanford University, Stanford, CA 94305, USA}

\maketitle 
\textbf{
  The identification of sources driving cosmic reionization, a major phase transition from neutral Hydrogen to ionized plasma around 600-800 Myr after the Big Bang\cite{dayal18,mason19,robertson22}, has been a matter of intense debate\cite{robertson22a}. Some models suggest that high ionizing emissivity and escape fractions ($f_{\rm esc}$) from quasars support their role in driving cosmic reionization\cite{madau15, mitra18}. Others propose that the high $f_{\rm esc}$\ values from bright galaxies generates sufficient ionizing radiation to drive this process\cite{naidu20}. Finally, a few studies suggest that the number density of faint galaxies, when combined with a stellar-mass-dependent model of ionizing efficiency and $f_{\rm esc}$\, can effectively dominate cosmic reionization\cite{finkelstein19,dayal20}. However, so far, low-mass galaxies have eluded comprehensive spectroscopic studies owing to their extreme faintness. Here we report an analysis of eight ultra-faint galaxies (in a very small field) during the epoch of reionization with absolute magnitudes between $M_{\rm UV}$ $\sim -17$ to $-15$ mag (down to 0.005 $L^{\star}$ \cite{finkelstein15,bouwens15}). We find that faint galaxies during the Universe's first billion years produce ionizing photons with log($\xi_{\mathrm{ion}}$/ Hz erg$^{-1}$) =$25.80\pm 0.14$, a factor of 4 higher than commonly assumed values\cite{robertson15}. If this field is representative of the large scale distribution of faint galaxies, the rate of ionizing photons exceeds that needed for reionization, even for escape fractions of order five per cent.
}
\bigskip

\noindent We combine ultra-deep {\em JWST}\ imaging data with ancillary {\em Hubble Space Telescope} ({\em HST}) imaging of the gravitational lensing cluster Abell~2744 (A2744 hereafter) in order to photometrically select extremely faint galaxy candidates in the epoch of reionization. A crucial component of the present study is the use of strong gravitational lensing to amplify the intrinsically-faint flux of distant sources. An accurate estimate of the magnification factor is required to retrieve the intrinsic luminosity of sources. This step relies on a good knowledge of the total mass distribution in the galaxy cluster. Here we use the most recent lensing model ({\tt v1.1}) published for the UNCOVER survey. The magnification factors for our galaxy sample range from $\mu \sim 2$ to $\mu \sim 27$. The values are reported in Table 1, together with 1$\sigma$ uncertainties. The second part of the UNCOVER program consists of ultra-deep follow-up spectroscopy with the NIRSpec instrument. We used the Multi-Shutter Assembly to obtain multi-object spectroscopy in 7 pointings, totaling an exposure time ranging from 2.7 to 17.4 hours. Figure 1 shows the position of these sources in the A2744 field with the associated regions of high magnification and the configuration of the NIRSpec slits. Simultaneous spectral fits to the continuum and the emission lines provide estimates of the spectroscopic redshifts of these sources, which lie between $z\sim6$ and $z\sim7.7$ (cf. Table 1). The spectral extraction and fitting procedure are discussed in the Methods section.

\begin{figure*}
    \centering
    \includegraphics[width=12cm]{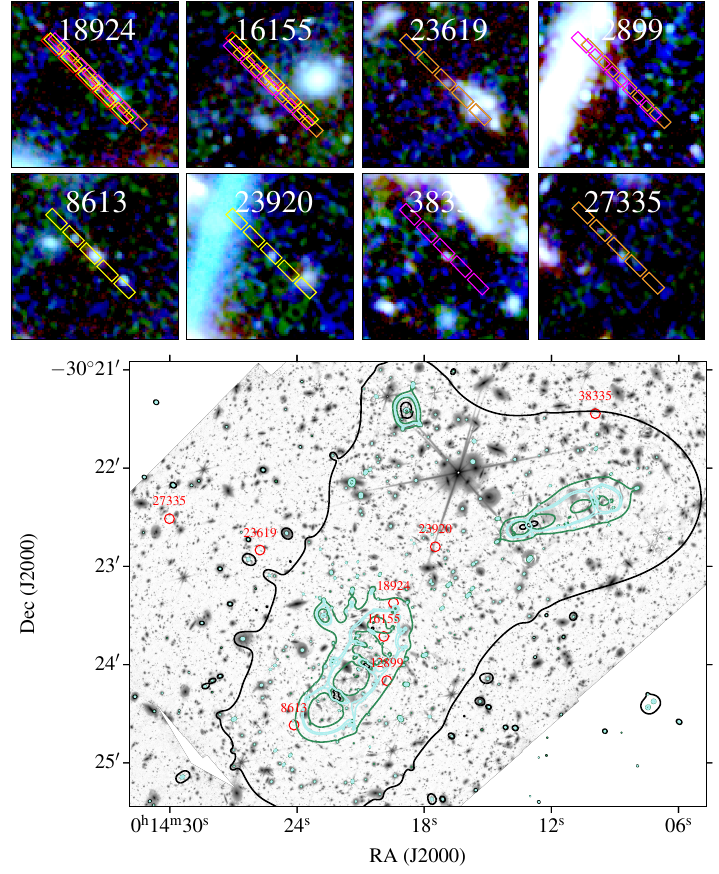}
    \caption{{\bf Layout of the ultra-faint galaxies identified in A2744 cluster field.} The figure shows a JWST LW filters stack (F277W+F356W+F444W) with magnification regions (at $z=7$) marked with black contours ($\mu >2$), green ($\mu >10$), and cyan ($\mu >100$), which are derived from the latest lensing model\cite{furtak23b}. The position of each source is marked with a red circle. Two of the sources (12899, 16155) are predicted to be multiply imaged by the lens model, but only the marked image of each system was targeted spectroscopically. On the top of the image, we show an RGB image of each source and the positions of each NIRSpec slitlet on top of the target.}
    \label{fig:enter-label}
\end{figure*}

\noindent Thanks to the gravitational magnification, we measure extremely faint line fluxes down to $f=5\times 10^{-21}$ erg~s$^{-1}$~cm$^{-2}$. We also derive intrinsic absolute magnitudes as faint as $M_{\rm UV}$~$\lesssim -15.34$ mag, which is nearly 2 magnitudes fainter than the faintest galaxies discovered in {\em JWST}\ spectroscopic surveys \cite{bunker23b,roberts-borsani23,mascia23} at the epoch of reionization so far (see Figure 2). Assuming a characteristic magnitude of $M_{UV}^{\star}=-21.15$ mag at $z=7$, these galaxies are as faint as 0.005$L^{\star}$. In light of the steep faint-end slope of the galaxy UV luminosity function at $z>6$ \cite{finkelstein15,ishigaki18}, such galaxies likely provide the bulk of the UV radiation at the epoch of reionization \cite{dayal20,robertson22}. In order to infer the stellar populations of these systems, we perform joint spectro-photometric SED fits using the \texttt{Bagpipes} software package (\texttt{Bayesian Analysis of Galaxies for Physical Inference and Parameter EStimation}). Accounting for the magnification, we derive extremely-low stellar masses between log(M$_{\star}$/M$_{\odot}$) = $5.88_{-0.08}^{+0.13}$ and log(M$_{\star}$/M$_{\odot}$) = $7.12_{-0.08}^{+0.07}$. Our results also clearly show that these galaxies harbor very young stellar populations, with stellar ages mostly around a few million years (cf. Table 2). This picture is also supported by their blue UV continuum slopes, derived from our SED-fitting, in the range $\beta=[-2.07, -2.53]$. Indeed, these values are generally indicative of a young massive stellar population and low dust attenuation.

\begin{figure*}
    \centering
    \includegraphics[width=8cm]{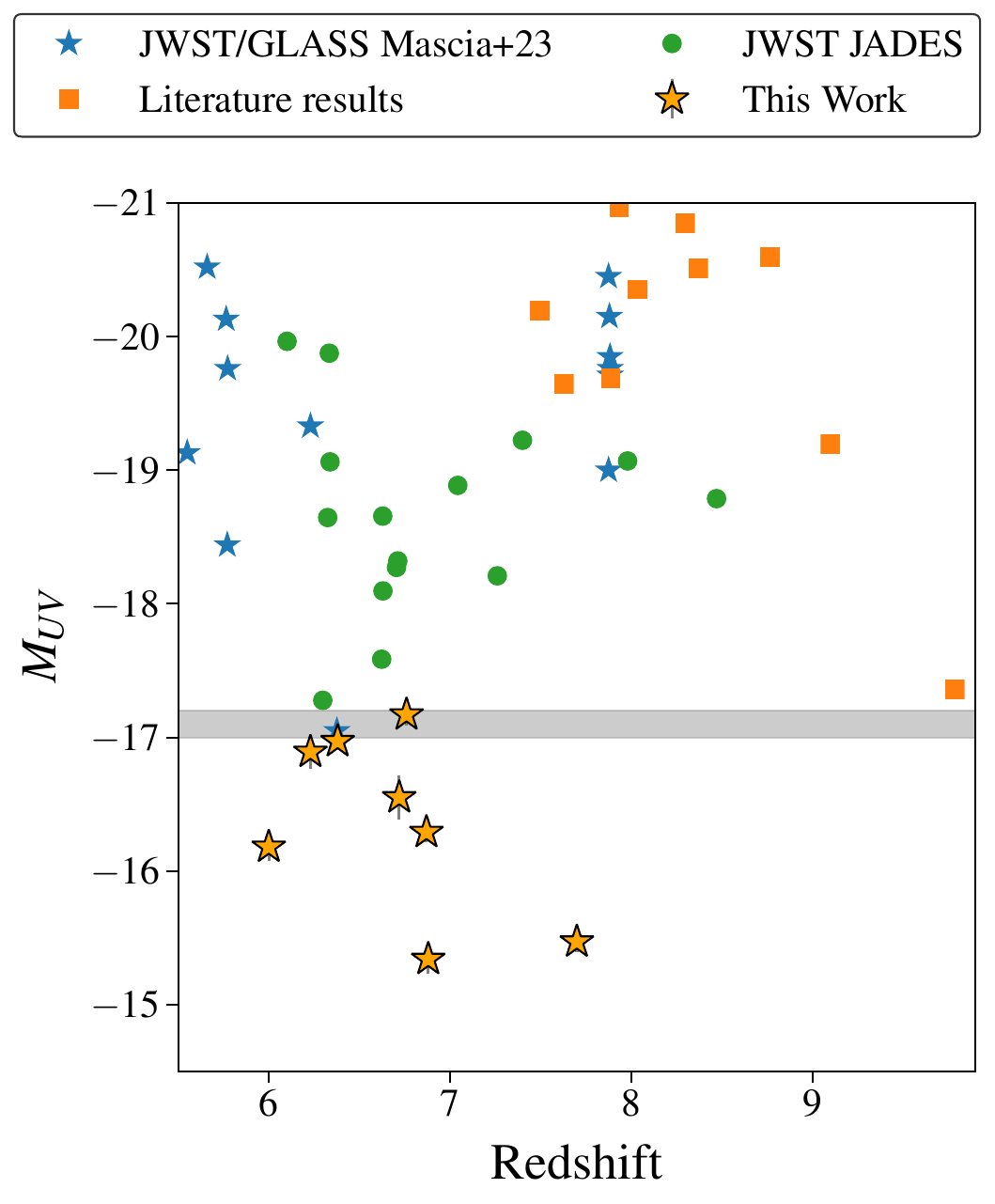}
    \caption{ {\bf Spectroscopic observations of the faintest galaxies during the epoch of reionization.} Various literature results from ground-based, {\em HST}\ and {\em JWST}\ observations are shown with orange squares \cite{roberts-borsani23}. The blue stars represent the spectroscopic sample of the {\em JWST}/GLASS survey presented in \cite{mascia23}. The green circles are derived from the latest data release of the deep spectroscopic observations of the {\em JWST}/JADES program \cite{bunker23b}. The horizontal gray line denotes the limit of the deepest {\em JWST}\ spectroscopic programs.}
    \label{fig:enter-label}
\end{figure*}

\noindent The ability of galaxies to reionize the Universe depends on their production of ionizing photon density per unit of time and the fraction of this radiation that escapes to ionize the intergalactic neutral gas. This quantity can be summarized by the following relation: $\dot{n}_{\rm ion} = f_{\rm esc}$ $\xi_{\mathrm{ion}}$\ $\rho_{UV}$, where $\rho_{UV}$ is the non-ionizing UV luminosity density at 1500 \AA, $\xi_{\mathrm{ion}}$\ is the ionizing photon production efficiency which represents the number of ionizing photons (Lyman continuum photons, LyC) per unit UV luminosity density, and $f_{\rm esc}$\ is the fraction of this LyC radiation that escapes the galaxy to ionize the IGM\cite{robertson22}. It is now well-established that faint galaxies ($M_{\rm UV}$$>-18$) are the dominant source of UV radiation during the reionization period \cite{atek15b,finkelstein15,bouwens17b}, although recent {\em JWST}\ observations are unveiling a number of faint active galactic nuclei (AGNs) at $3<z<7$\cite{matthee23b}. The ionizing properties of these galaxies however are virtually unknown. For example, measurements of their ionizing efficiency $\xi_{\mathrm{ion}}$\ have proven challenging even with the deepest {\em JWST}\ spectroscopic surveys, which are limited to galaxies brighter than $M_{\rm UV}$$\sim-18$\cite{fujimoto23}. We report direct spectroscopic measurements of the ionizing efficiency $\xi_{\mathrm{ion}}$\ of the faint population of galaxies during the epoch of reionization.

\noindent In Figure 3 we present our $\xi_{\mathrm{ion}}$\ measurements based on the ratio of H$\alpha$\ recombination line, which is powered by reprocessed ionizing radiation, and the non-ionizing UV luminosity. The most striking result is the high value of log($\xi_{\mathrm{ion}}$/ Hz erg$^{-1}$) =$25.80\pm 0.14$ observed in faint ($M_{\rm UV}$$>-16.5$) galaxies, compared to the canonical value of log($\xi_{\mathrm{ion}}$/ Hz erg$^{-1}$ =25.2)\cite{robertson15} commonly assumed in reionization models, or previous studies at this epoch. For example, the measured efficiency in a population of Ly$\alpha$ emitters, which are thought to have larger ionization radiation than the average galaxy population, is around log($\xi_{\mathrm{ion}}$/ Hz erg$^{-1}$) =25.4\cite{simmonds23}. Our measured value is consistent with the maximum values predicted by the BPASS \cite{stanway18} stellar population models for a dust-free galaxy with a constant star formation and a stellar age of less than 3 Myr and a 0.1$Z_{\odot}$ metallicity. Such a large ionizing efficiency in faint galaxies implies that modest values of $f_{\rm esc}$\ are sufficient for galaxies to reionize the Universe by $z=6$. Until now, most models of reionization needed to assume large values of $f_{\rm esc}$, typically around 20\%, to accommodate the relatively low ionizing photon emissivity observed in high$-z$ galaxies. Some models required lower $f_{\rm esc}$\ values with combinations of specific galaxy properties and a small contribution from AGNs \cite{finkelstein19}. Direct measurements of $f_{\rm esc}$\ $z=0-4$ analogs, reported typical or sample-averaged values below 10\% \cite{pahl21}, albeit with a large dispersion and higher values have been observed in individual objects. As we can see in Figure 3, a volume-averaged escape fraction as low as $f_{\rm esc}$$=5\%$ is sufficient for faint galaxies to maintain reionization. 

\begin{figure*}
    \centering
    \includegraphics[width=12cm]{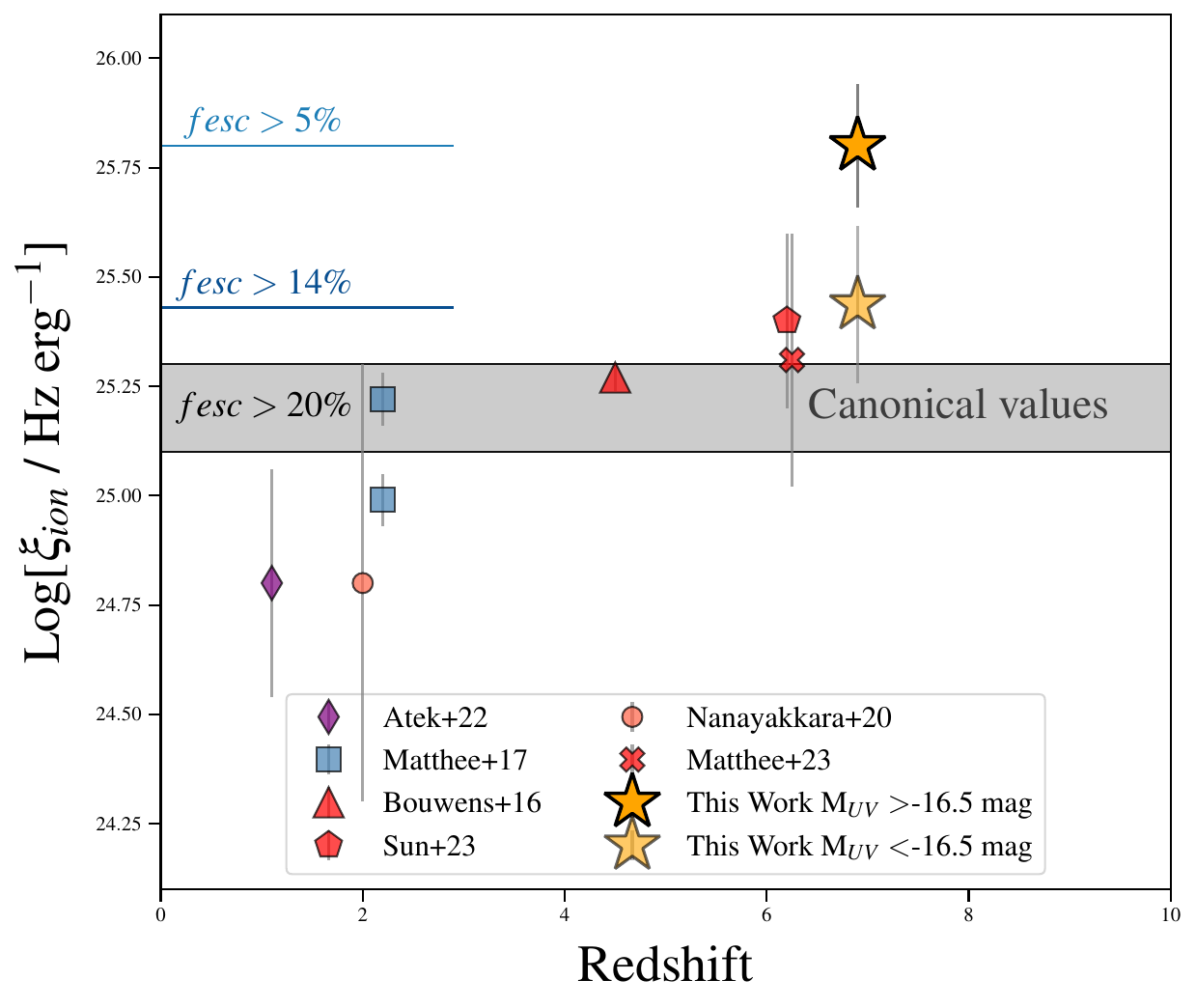}
    \caption{{\bf The ionizing photon production efficiency of faint galaxies during the epoch of reionization}. Our $\xi_{\mathrm{ion}}$\ measurements are marked with an orange star (light and dark shade for galaxies brighter and fainter than $M_{\rm UV}$$=-$16.5, respectively). The gray-shaded horizontal line represents the canonical values assumed when assessing the contribution of galaxies to reionization. Various literature results are also shown and listed in the Methods section. Assuming a fiducial UV luminosity density \cite{atek15b,finkelstein15}, we plot the minimum $f_{\rm esc}$\ required to maintain reionization at each given value of $\xi_{\mathrm{ion}}$ (horizontal lines). The error bars represent 1-$\sigma$ uncertainties.}
    \label{fig:enter-label}
\end{figure*}

\noindent To paint a complete picture of cosmic reionization by star-forming galaxies, we compute the spectroscopic UV luminosity function based on the present sample. We put spectroscopic constraints on the prevalence of ultra-faint galaxies. Albeit with a small sample size, our measurements provide a confirmation of the steep faint-end slope of the UV LF at $z\sim7$, in agreement with the photometric UV LF derived from HFF observations\cite{atek18}. By integrating this UV LF down to a faint-end limit of $M_{\rm UV}$$=-15$ mag, we determine a UV luminosity density of log($\rho_{UV}$ / erg s$^{-1}$ Mpc$^{-3}$)=26.22. Now combining these two quantities we can infer the total ionizing emissivity of galaxies for different values of $f_{\rm esc}$, accounting for the contribution of the ultra-faint population. The result is shown in Figure 4. Galaxies clearly produce enough ionizing photons to maintain reionization at $z\sim7$\cite{gnedin22}, assuming on average as little as 5\% of this radiation escape from the galaxies to heat the IGM. We can go a step further by indirectly estimating $f_{\rm esc}$\ using the UV continuum slope we measured for these galaxies. Specifically, we follow an approach pioneered by recent studies of nearby galaxies that calibrated indirect indicators of $f_{\rm esc}$\ of LyC emission. In particular, a strong correlation is observed between the observed $\beta_{obs}^{1550}$ slope and $f_{\rm esc}$ \cite{chisholm22}. Adopting the UV-slope $\beta$ (see Table 2) as a proxy for $f_{\rm esc}$, we infer escape fractions within $f_{\rm esc}$=[0.045, 0.16], well in the range of assumed values in Figure 4. Another indirect indicator of $f_{\rm esc}$\ that has been explored in recent studies is the star formation surface density $\Sigma_{\rm SFR}$ \cite{naidu22}. For these compact sources, we measure log($\Sigma_{\rm SFR}$/ M$_{\odot}$yr kpc$^{2}$) $\sim 0.2 - 2$. These values are commonly observed in LyC leakers \cite{naidu17,vanzella18} and also predictive of high $f_{\rm esc}$\ according to published best-fit relations of $f_{\rm esc}$-$\Sigma_{\rm SFR}$\cite{naidu20}.

\begin{figure*}
    \centering
    \includegraphics[width=10cm]{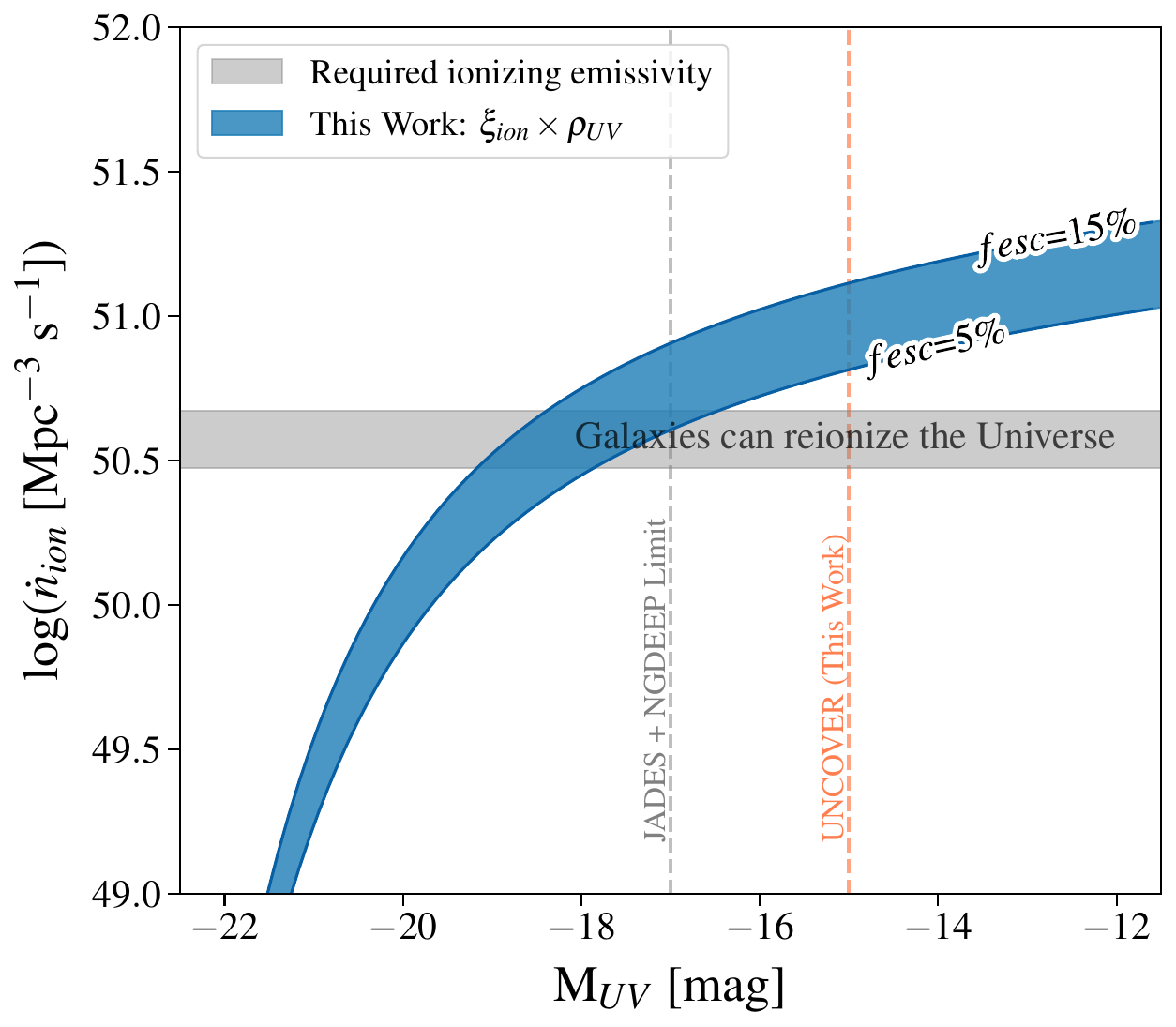}
    \caption{ {\bf The total ionizing emissivity of galaxies at $z\sim7$.} The total ionizing photon production rate density, derived from the prevalence and the ionizing efficiency of galaxies, as a function of the faint integration limit. The blue region delimits the two cases where $f_{\rm esc}$=5\% and $f_{\rm esc}$=15\%. The gray-shaded region is the threshold required to maintain the Universe ionized at $z=7$. The gray vertical line marks the magnitude limit of the deepest {\em JWST}\ spectroscopic surveys to date. The orange vertical line shows the limit probed by this work. At this luminosity, galaxies produce enough radiation to reionize the Universe.}
    \label{fig:enter-label}
\end{figure*}

\bigskip

\noindent We note that measurements of $\xi_{\mathrm{ion}}$\ can be significantly affected by dust attenuation. This concern also applies to our estimate of stochastic star-formation through the SFR(H$\alpha$)/SFR(UV) ratio. However, as indicated by the blue UV continuum slopes that we observe, we expect the dust content to be small in these galaxies. This assumption is also supported by the low Balmer decrement H$\alpha$/H$\beta$, for which we measure an average value of $3.3\pm0.5$. Therefore, dust attenuation should not affect significantly these quantities.  We note that we used indirect indicators, which come with a significant scatter, to estimate $f_{\rm esc}$, because direct measurements of LyC at the epoch of reionization are impossible. The stochastic nature of star formation in these low-mass galaxies also makes the $f_{\rm esc}$\ highly variable, since it mainly relies on stellar and supernovae feedback clearing the ISM for LyC escape \cite{trebitsch17}. However, on average, hydrodynamical simulations predict higher $f_{\rm esc}$\ in lower-mass galaxies\cite{ma20,yeh23}. Again, based on the ionizing photon production we estimated, modest values of $f_{\rm esc}$\ around 5\% are sufficient. We also note that our conclusions are based on observations obtained in one field, and are therefore not immune to field-to-field variations or environmental effects. For instance, the ionizing properties of faint galaxies can be impacted differently by reionization radiation if they reside in over-dense regions \cite{hutter21}. Additional observations in an independent field should provide further insights in that regard.

\pagebreak
\begin{table}[!h]
    \centering
    \caption{{\bf Summary of the sample properties.} The exposure time (4) corresponds to the total of all NIRSpec observations for each source. The magnification factors (5) are computed at the spectroscopic redshift of the source using the most recent UNCOVER lensing model. We also added systematic uncertainties derived from a comparison with an independent lensing model \cite{bergamini23}. The photometric redshift (6) is measured with the {\tt Eazy} software. The spectroscopic redshift (7) is measured from the best {\tt msaexp} fit. The typical best-fit error is $\sigma_{zspec}=0.01$.  Absolute magnitude $M_{\rm UV}$\ (8) is measured in the rest-frame UV using the observed magnitude derived from the UNCOVER photometric catalog corrected for magnification.}
    \begin{tabular}{l c c c c c  c c}
    \hline
       Source  & RA & Dec & Exptime &$\mu$ & $z_{phot}$ & $z_{spec}$ & $M_{\rm UV}$   \\
         & J2000 & J2000 & hours &&  & &AB    \\
       \hline 
    18924   &3.581044&-30.389561& 17.4	&$26.6\pm7.1$   &$7.9_{0.3}^{0.2}$  & 7.70 & $-15.47\pm0.08$ \\
    16155	&3.582953&-30.395232& 17.4	&$11.1\pm3.8$	&$6.7_{0.1}^{0.1}$  & 6.87 & $-16.29\pm0.08$   \\
    23920	&3.572830&-30.380026& 3.7	&$3.3\pm0.1 $	&$6.4_{4.9}^{0.4}$  & 6.00 & $-16.18\pm0.10$   \\
    12899	&3.582353&-30.402732& 10.2	&$13.9\pm0.9$	&$6.6_{0.2}^{0.2}$  & 6.88 & $-15.34\pm0.11$  \\
    8613	&3.600602&-30.410271& 2.7	&$9.3\pm0.6 $   &$6.5_{0.1}^{0.1}$  & 6.38  & $-16.97\pm0.04$  \\
    23619   &3.607272&-30.380578& 7.5   &$1.8\pm0.2 $   &$6.7_{0.3}^{0.2}$  & 6.72  & $-16.55\pm0.16$  \\
    38335   &3.541383&-30.357435& 2.7   &$2.3\pm0.2 $   &$6.4_{1.8}^{2.2}$  & 6.23  & $-16.89\pm0.13$  \\
    27335   &3.625081&-30.375261& 7.5   &$1.4\pm0.1 $   &$6.9_{0.1}^{0.5}$  & 6.76  & $-17.17\pm0.08$  \\
      \hline
    \end{tabular}
    \label{tab:sample}
\end{table}
\pagebreak
\begin{table}[!ht]
    \centering
    \caption{{\bf Summary of the physical properties of the sample derived from SED fitting with {\tt Bagpipes}.} For each source (1), the median posterior and associated uncertainties from the best-fit models are given for: the stellar mass (2), the star formation rate SFR (4), the UV continuum slope $\beta$ (5), the half-mass age (6). The oxygen abundance computed from strong optical lines is also reported (7). The star formation rate derived from the H$\alpha$\ emission is also reported in column (3)}
    \begin{tabular}{l c c c c c c}
    \hline 
     Source  &log(M$_{\star}$/M$_{\odot}$) & SFR$_{{\rm H}\alpha}$  & SFR$_{\rm UV}$   & $\beta$ & $t_{\mathrm{50}}$ & 12+Log(O/H)\\
             &                  &  M$_{\odot}$yr        &  M$_{\odot}$yr &       &         Myr             &            \\
     \hline 
18924 & $5.88_{-0.08}^{+0.13}$ & $0.33\pm0.02$  & $0.01_{-0.07}^{+0.14}$ & $-2.39_{-0.10}^{+0.12}$ & $2.23_{-0.85}^{+0.68}$ & $6.95\pm0.15$ \\ 
16155 & $6.61_{-0.06}^{+0.07}$ & $0.92\pm0.04$  & $0.04_{-0.06}^{+0.08}$ & $-2.09_{-0.08}^{+0.07}$ & $3.96_{-0.66}^{+0.92}$ & $7.01\pm0.19$ \\ 
23920 & $6.30_{-0.03}^{+0.03}$ & $1.32\pm0.04$  & $0.02_{-0.03}^{+0.03}$ & $-2.45_{-0.03}^{+0.03}$ & $1.12_{-0.11}^{+0.32}$ & $6.84\pm0.06$ \\ 
12899 & $6.54_{-0.19}^{+0.14}$ & $0.49\pm0.02$  & $0.04_{-0.15}^{+0.12}$ & $-2.51_{-0.07}^{+0.09}$ & $28.66_{-11.98}^{+15.51}$ & $6.70\pm0.15$ \\ 
 8613 & $7.12_{-0.08}^{+0.07}$ & $0.78\pm0.07$  & $0.16_{-0.07}^{+0.08}$ & $-2.53_{-0.03}^{+0.04}$ & $25.73_{-6.33}^{+6.47}$ & $6.97\pm0.18$ \\ 
23619 & $6.57_{-0.06}^{+0.10}$ & $0.85\pm0.07$  & $0.04_{-0.05}^{+0.11}$ & $-2.51_{-0.07}^{+0.13}$ & $1.08_{-0.07}^{+0.22}$ & $7.19\pm0.20$ \\ 
38335 & $6.83_{-0.20}^{+0.25}$ & $1.00\pm0.16$  & $0.07_{-0.15}^{+0.34}$ & $-2.07_{-0.24}^{+0.29}$ & $6.45_{-2.29}^{+4.39}$ & $7.46\pm0.32$ \\ 
27335 & $6.73_{-0.08}^{+0.15}$ & $0.73\pm0.10$  & $0.05_{-0.07}^{+0.17}$ & $-2.35_{-0.11}^{+0.22}$ & $1.56_{-0.52}^{+1.33}$ & $6.99\pm0.18$ \\ 
  \hline 
    \end{tabular}
    \label{tab:props}
\end{table}
\pagebreak


\section*{Methods}
\label{sec:methods}

 Throughout the paper, we use AB magnitudes \cite{oke83} and a standard cosmology with H${_0} =70$ km s$^{-1}$ Mpc$^{-1}$, $\Omega_{\Lambda}=0.7$, and $\Omega_m=0.3$.

\subsection*{Observations and sample selection} 

The UNCOVER dataset consists of both imaging and spectroscopic observations of the lensing cluster A2744. The imaging observations and data reduction are described in detail in the survey and catalog papers \cite{bezanson22,weaver23}. Here, we briefly summarize the imaging and photometric products used in the present paper. HST imaging consists of 7 broadband filters (F435W, F606W, F814W, F105W, F125W, F140W, F160W). The NIRCam\cite{rieke23} images include short-wavelength (SW) broadband filters (F115W, F150W, F200W), long-wavelength (LW) broadbands (F277W, F356W, F444W), and one medium-band filter (F410M). Data were processed, and drizzled into 0.04 arcsec pix$^{-1}$ mosaics using the Grism redshift and line analysis software for space-based spectroscopy ({\texttt Grizli}; v1.6.0.dev99)\cite{brammer19}. In terms of ancillary data, the HFF program \cite{lotz17} has obtained deep optical and NIR observations of the core area of A2744 with the Advanced Camera for Surveys (435W, F606W, F814W), and Wide-Field Camera Three (F105W, F125W, F140W, F160W). A wider area around the cluster has also been covered by the BUFFALO program \cite{steinhardt20} in almost identical broadband filters (without F435W and F140W). All {\em HST}\ observations were drizzled to the same pixel scale and the same orientation as the NIRCam mosaics. 

 \noindent The second part of the UNCOVER program consists of ultra-deep follow-up spectroscopy with the NIRSpec instrument \cite{jacobsen22}. Data were obtained between July 31st and August 2nd 2023. Observations use the Prism mode and the Multi-Shutter Assembly \cite{ferruit22} of NIRSpec to observe more than 650 targets. In order to optimize background subtraction, each target has been observed with a 3-slitlet nodding strategy. Observations were split into 7 pointings, with important overlap at the center, providing total on-target exposure times ranging from $\sim 2.7h$ to $\sim 17.4h$. The spectral resolution is wavelength-dependent and varies between $R\sim30$ to $R\sim300$ over the full wavelength range $\lambda \sim 0.6-5.3 \mu m$. Data were reduced using the JWST/NIRSpec analysis software {\tt msaexp} version 0.6.10. The processing is based on level 2 MAST products, using the CRDS context file {\tt jwst\_1100.pmap}. The software performs basic reduction steps, including flat-field, bias, 1/f noise and snowballs correction, wavelength and photometric calibrations of individual exposure frames \cite{heintz23}. The extraction of 1D spectra from individual exposures is operated on inverse-weighted stack of 2D spectrum in the dispersion direction, following an optimal extraction procedure \cite{horne86}. Then the software fits a Gaussian profile along the cross-dispersion direction to define the 1D extraction aperture. Finally, we compute the final deep 1D spectrum by inverse-variance stacking the individual spectra. In order to account for slit loss effects, we apply a wavelength- (broadband-) dependent correction factor to re-scale the 1D spectrum to the observed NIRCam aperture photometry. We show an example of the imaging and spectroscopic data in Figure 5. A clear Lyman-break at rest-frame wavelength $\lambda_{\rm rest}=1216$ \AA\ is observed, together with multiple strong emission lines, including H$\alpha$+[N{\sc ii}], [O{\sc iii}]$\lambda\lambda4960, 5008$, H$\beta$, H$\gamma$, and [O{\sc ii}]$\lambda3727$

\noindent  The selection of our sample combines several criteria  to constrain the photometric redshifts of the sources. First, we applied a color-color selection, based on a flux-dropout in the {\em HST}\ optical filters caused by rest-frame Lyman$-\alpha$ absorption by residual intergalactic Hydrogen gas. This selection consolidates most of the sources identified in the Hubble Frontier Fields (HFF) data \cite{atek18,bouwens22c} at $6<z<9$. Second, we performed spectral energy distribution (SED) fitting with the {\tt Eazy} \cite{brammer22} software to estimate photometric redshifts, assuming a flat luminosity prior and the {\tt corr\_sfhz} library of stellar population templates. The allowed redshift range was set to $0.01<z<20$. The sources have been selected to have best-fit photometric solutions lying $6<z<9$ at the heart of the epoch of reionization. The final selection was then performed according to the intrinsic luminosity, combining high magnifications ($\mu \gtrsim 2$) and faint observed luminosities in F150W, resulting in intrinsic absolute UV magnitudes of order $M_{\rm UV}$$\gtrsim -17$.

\begin{figure*}
    \centering
    \includegraphics[width=12cm]{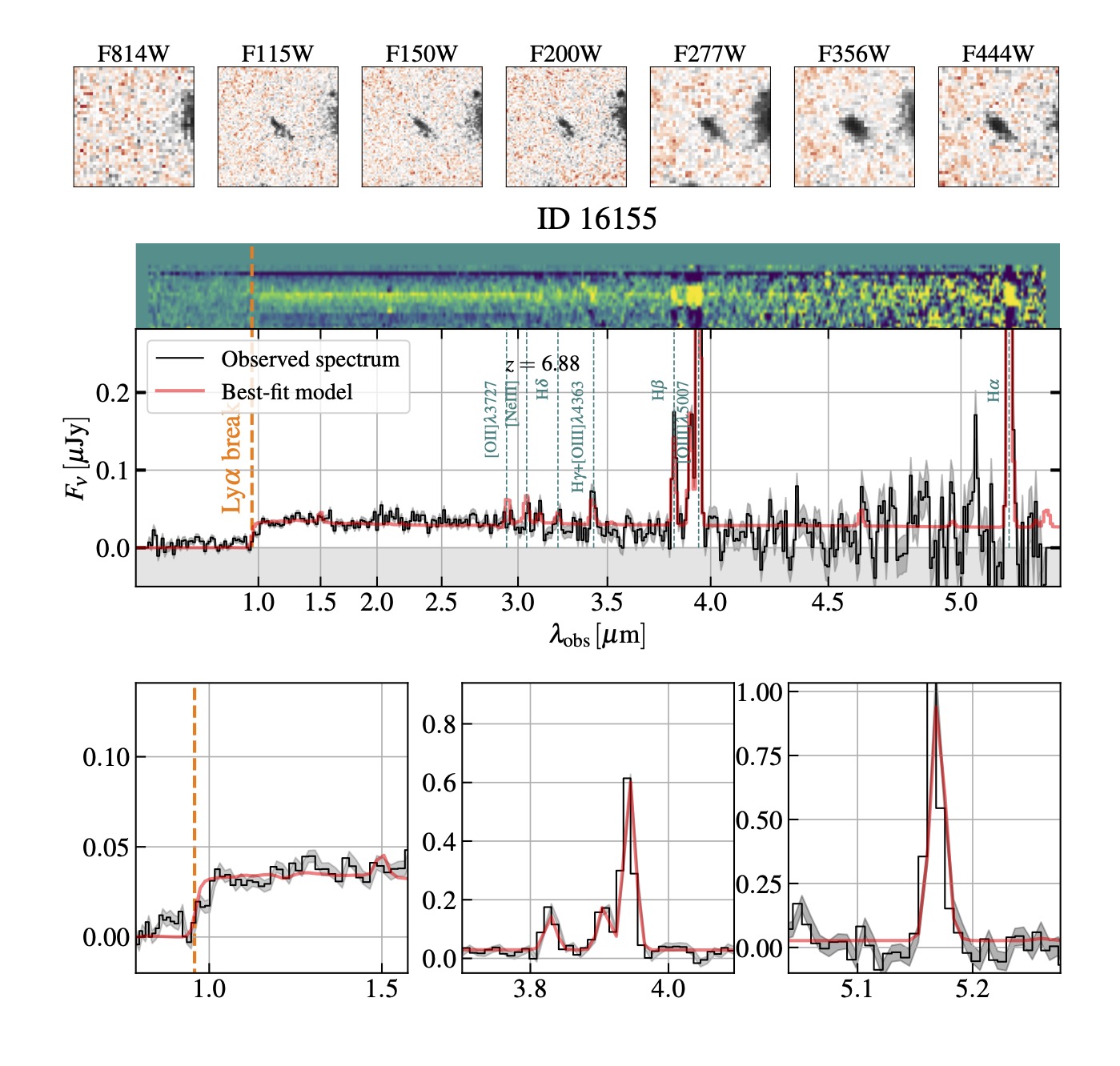}
    \caption{{\bf UNCOVER JWST data for galaxy 16155 at $z_\mathrm{spec}=6.88$}. The top panels show image cutouts in seven different filters at increasing wavelength including ancillary HST/ACS data in F814W, and UNCOVER JWST imaging in F115W, F150W, F200W, F277W, F356W, and F444W bands (left to right). The central panel shows the UNCOVER NIRSpec data, with the 2D spectrum on top of the 1D optimally extracted spectrum (black with gray 1-$\sigma$ uncertainty ranges). The red lines show the best-fit {\tt msaexp} template spectrum. The observed-frame wavelengths of key emission lines are indicated as vertical dashed lines. The bottom panels show a zoomed in version of three different parts of the spectrum around the Ly$\alpha$ break (left), around the [O{\sc iii}]+H$\beta$ emission lines (middle) and the H$\alpha$ line (right).}
    \label{fig:enter-label}
\end{figure*}

\subsection*{Spectral fitting}

In order to determine the spectroscopic redshift of the sources, we fit spectral templates using {\tt msaexp}, which based on the SED fitting software {\tt Eazy}\cite{Brammer2008}. The code combine a set of templates to fit simultaneously the continuum, including the Lyman break caused by the IGM absorption, and the emission lines. For our analysis, we adopt the \texttt{corr\_sfhz\_13} template library which include redshift-dependent SFHs, which are known to perform better than the default {\tt fsps\_full} library in recovering the true redshift \cite{weaver23}. We search for the best-fit solution over the redshift interval $0<z<15$. The spectroscopic redshifts are reported in Table 2. Once the best-fit redshift is found, we refit the spectra, fixing the redshift to $z_{\mathrm{spec}}$, with a set of spline functions to measure the continuum and Gaussians in order to measure the emission line fluxes. Examples of the best-fit model plotted over the observed spectrum are presented in Figure 1.

\subsection*{Strong lensing}
We use version \texttt{v1.1} of the UNCOVER lensing model\cite{furtak23b}, which is publicly available in the latest UNCOVER data release DR-1. The model is based on the parametric approach by Zitrin et al. \cite{zitrin15b} which has been re-written to be fully analytic, i.e. not limited by a grid-resolution \cite{pascale22,furtak23b}. The UNCOVER lens model of A2744 was constructed on a wealth of ground- and space-based data, including deep {\em HST}\ and {\em JWST}\ imaging, and \textit{Multi Unit Spectroscopic Explorer} \cite{bacon10} (MUSE) spectroscopic redshifts of both cluster members and multiple images \cite{mahler18,richard21,bergamini23,bergamini23b}. It comprises 421 cluster member galaxies identified in the $\sim45\,\mathrm{arcmin}^2$ UNCOVER field-of-view and five smooth cluster-scale dark matter (DM) halos. The model is constrained with 141 multiple images belonging to 48 sources and achieves an image reproduction RMS of $\Delta_{\mathrm{RMS}}=0.51''$ in the lens plane. Thanks to the massive cluster substructures identified with UNCOVER \cite{furtak23b}, the critical area of the cluster is $1.5\times$ larger than inferred from HFF data and the total source plane area with $\mu>4$ of $\sim4\,\mathrm{arcmin}^2$ for a source at redshift $z_{\mathrm{s}}=6$. The model uncertainties on the amplification values are derived from a Markov Chain Monte Carlo (MCMC) procedure within the modeling code {\tt Zitrin-analytic} \cite{zitrin15,furtak23a}. The temperatures of the MCMC are chosen to reflect typical systematics inherent to parametric lens modeling techniques \cite{furtak23a}. In order to better estimate systematics uncertainties inherent to models, we used an independent mass model for A2744 \cite{bergamini23} for comparison. The magnification factors are in good agreement within 1-$\sigma$ uncertainties, except for two objects that have a difference at the level of 1.5 and 2-$\sigma$. Moreover, the statistical uncertainties derived from each model are of the same order. We have incorporated these systematic uncertainties in the quoted errors on the amplification factors.

\subsection*{Physical properties}

\subsubsection*{Bagpipes}

We infer global physical properties from SED-fitting using the \texttt{Bagpipes} software package \cite{Carnall2018,Carnall2019:VANDELS}). Before fitting, all models are convolved with the NIRSpec/Prism instrumental resolution curve provided by the Space Telescope Science Institute (\texttt{jwst\_nirspec\_Prism\_disp.fits}), assuming that the flight performance is 1.3 times better than stated, and is consistent with an earlier work where a factor 1/0.7 is introduced for modeling $z>10$ galaxies \cite{curtis-lake23}. Additionally, we fit with a wavelength-independent velocity smoothing ($0< $ log$(v_\mathrm{smooth})<3.3$) as a nuisance parameter. We adapt the following model grid: \cite{bc03} stellar population models, the MILES spectral library \cite{Sanchez-Blazquez2006,Falcon-Barroso2011}, CLOUDY nebular emission models \cite{Ferland:2017}, and \cite{Charlot2000} dust model (with $0<A_{v}<5$ and $0.3<n<2.5$ as free parameters). The stellar and gas phase metallicity are tied to the same value, and also included as a free parameter in the range $-2<\mathrm{log(Z/Z_\odot)}<0.3$. The ionization parameter is also left free in the range of $-3.5<\mathrm{log(U)}<-1.0$. We parameterize the star formation history as a delayed-$\tau$ model (SFR$\propto^{-t/\tau}$, which can flexibly produce rising or falling star formation histories for this range of $\tau$ at the redshift of the sample), with the age ($-3<$log(age)$<0.48$) and $\tau$ (0.01$<\tau$<5) as free parameters. This parameterization has been shown to reliably recover star formation rates and the mass formed in recent star formation, but is potentially susceptible to under-estimating stellar masses due to outshining by the youngest stellar population \cite{Papovich2023}. Redshift is restricted to vary in a narrow range around the best-fitting spectroscopic redshifts ($\pm$0.1). We fit for a polynomial calibration vector of order 2 after applying a wavelength-independent calibration to scale the normalization of the spectrum to the photometry. The \texttt{Bagpipes} white noise model is used to allow for underestimated errors up to a factor of 10. A signal-to-noise ceiling of 20 is imposed on both our photometry and spectroscopy to account for systematic issues with the flux calibration. Sampling is performed via \texttt{PyMultinest} \cite{Buchner:2014, Feroz:2019}, with the default \texttt{Bagpipes} convergence criteria. The most important physical properties derived from this procedure are presented in Table 2, and example fits and posteriors are presented in Figure 6. 

\noindent  In addition to the star-formation rate (SFR) derived from SED-fitting, we also compute the SFR based on H$\alpha$\ or H$\beta$\ recombination line. The H$\alpha$\ indicator traces massive short-lived stars on a timescale of a few Myr, whereas the UV emission indicates an SFR averaged on a longer timescale up to a few 100 Myr. We report high values for the SFR(H$\alpha$)/SFR(UV) ratio in the range [5-60], indicating recent bursts of star-formation in these young systems. This is in line with their specific star-formation rates (sSFR). Given their low stellar masses, these sources have log(sSFR(H$\alpha$) / yr$^{-1}$)=[$-7.4, -6.2$], which means that they can double their stellar mass within 2 to 20 Myr.

\begin{figure*}
    \centering
    \includegraphics[width=11cm]{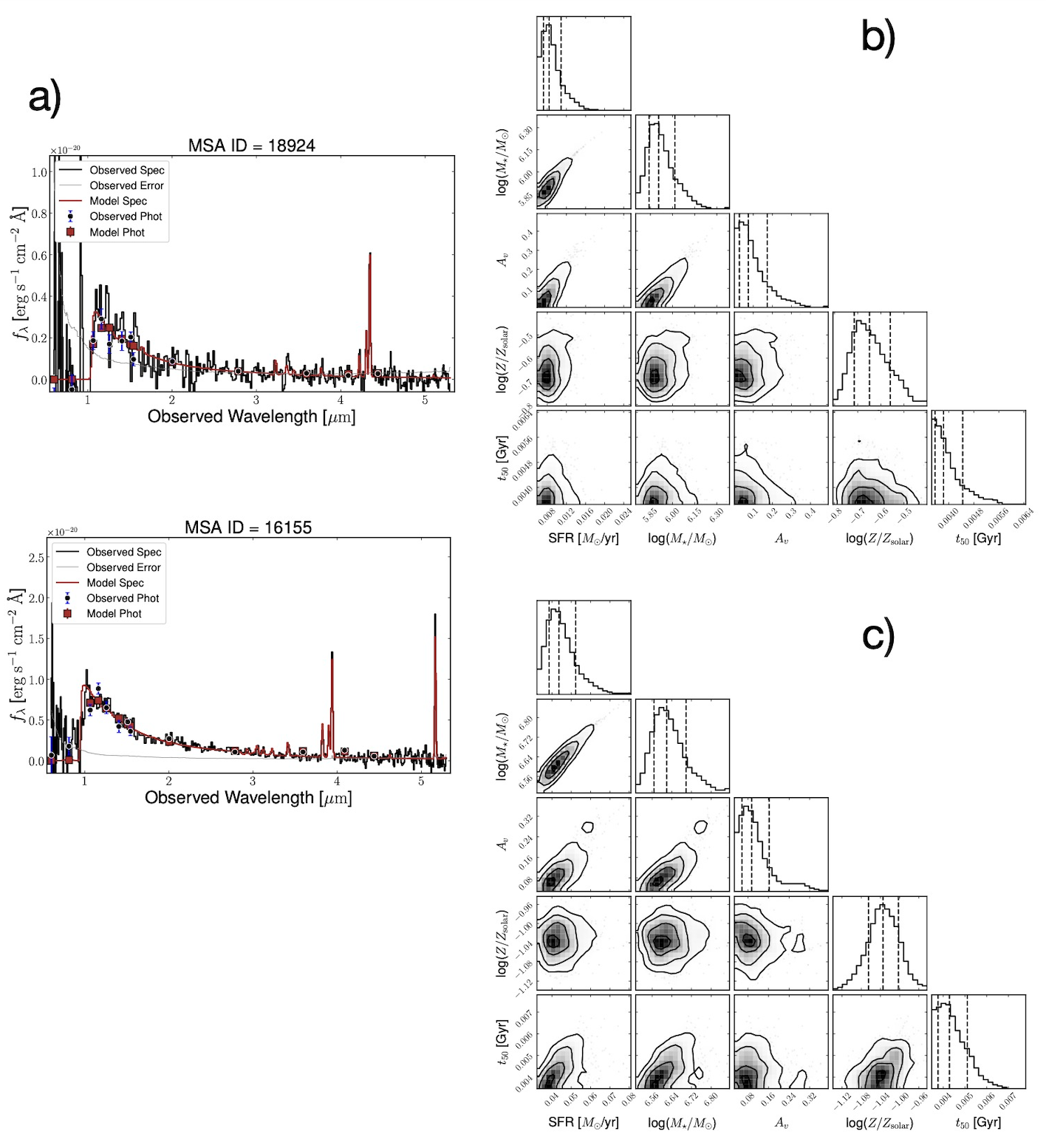}
    \caption{ {\bf Stellar population simultaneous fitting to the NIRSpec spectra and NIRCam photometry}. {\bf Panel a:} Two representative sources (IDs 18924 and 16155) are shown. The best-fit {\tt Bagpipes} model (red curve) is plotted over the observed NIRSpec spectrum (black curve), together with the error spectrum (gray curve). The NIRCam photometric measurements are represented with black points with their associated 1-sigma uncertainties. {\bf Panel b:} Posterior distribution function for the main physical properties of ID 16155. When relevant, the parameters are corrected for magnification. {\bf Panel c:} Same as panel b, for source ID 18924. }
    \label{fig:enter-label}
\end{figure*}

\subsubsection*{\texttt{BEAGLE}}

We run an additional spectral fit with the \texttt{BayEsian Analysis of GaLaxy sEds}\cite{chevallard16} tool (\texttt{BEAGLE}) on the magnification-corrected spectra. \texttt{BEAGLE} uses the latest version of the Bruzual \& Charlot stellar population synthesis models\cite{Bruzual_Charlot_2003} and nebular emission templates computed with \texttt{CLOUDY}\cite{ferland13,gutkin16}. We then assume a Chabrier\cite{chabrier03} initial stellar mass function (IMF), an SMC dust attenuation law\cite{pei92}, the latest Inoue et al. analytic IGM attenuation models\cite{inoue14}, and a delayed exponential SFH as for our \texttt{Bagpipes} fit. All other parameters are left free to vary with uniform or log-uniform priors: stellar mass $\log(M/\mathrm{M}_{\odot})\in[4,10]$, current (10~Myr) SFR $\log(\psi/\mathrm{M}_{\odot}\,\mathrm{yr}^{-1})\in[-2,4]$, maximum stellar age $\log(t_{\mathrm{age}}/\mathrm{yr})\in[6,t_{\mathrm{universe}}]$, star-formation e-folding time $\log(\tau/\mathrm{yr})\in[5.5,9.5]$, stellar metallicity $\log(Z/\mathrm{Z}_{\odot})\in[-2.2,-0.3]$, effective \textit{V}-band dust attenuation optical depth $\hat{\tau}_V\in[0,3]$, effective galaxy-wide ionization parameter $\log~U\in[-4,-1]$, gas-phase metallicity $\log(Z_{\mathrm{gas}}/\mathrm{Z}_{\odot})\in[-2.2,-0.3]$, and dust-to-metal mass ratio $\xi_{\mathrm{d}}\in[0.1,0.5]$. The posterior distribution of the physical properties that we derive with \texttt{BEAGLE} agree well with the \texttt{Bagpipes} results presented in Table 2.

\subsection*{Contribution of galaxies to reionization}

Using the present spectroscopically-confirmed sample of ultra-faint galaxies, we have the opportunity to put constraints on the UV luminosity function. We first describe the selection procedure of our sample and associated biases. The original sample has been selected in HFF studies\cite{atek18,bouwens22c}, based on {\em HST}\ observation. The selection of the original photometric sample is based on Lyman break criteria, which identify the dropout due to continuum absorption by the neutral IGM blueward of Ly$\alpha$. The photometric redshift through SED fitting were only measured to refine the redshift solution. In addition, three sources were selected from the UNCOVER imaging data based on their photometric redshifts. Regarding these three sources, strong emission lines tend to help put stronger constraints on the photometric redshift estimates, resulting in narrower best-fit solutions, which could favor strong-line emitters in the sample selection \cite{roberts-borsani16}. For the selection of this spectroscopic sample, we primarily focus on faint intrinsic magnitudes, typically $M_{\rm UV}$$\gtrsim-17$ mag, as can be seen in Figure 2. Their apparent magnitude ranges from $m_{F150W} =27.4$ to 29.7 AB mag. While there is an intentional bias to select intrinsically faint galaxies in this study, this is less the case regarding observed magnitudes. Finally, during the Multi-Shutter Assembly design, we assigned equal weights to all galaxies. Therefore, the only bias introduced here is the optimization of the number of sources that are included in one mask configuration. Therefore, galaxies that did not make it to the final sample were simply excluded for mask optimization reasons.

\noindent First, we compute an initial UV LF based on the present sample binned in four magnitude bins and a survey volume, which depends on the original selection of the source. For the five HFF sources, we use the source plane effective volume\cite{atek18} as a function of the magnitude bin. We rebin the original HFF sample to match the new magnitude bins. A scaling factor is then applied to the LF points to match the HFF completeness-corrected counts. Finally, we apply a correction factor based on the success rate of the spectroscopic confirmation. For the three sources outside the HFF area, we recompute the source plane effective survey volume using the new lensing model and assuming a similar completeness function across the field. We perform the same exercise of rescaling the number counts in each magnitude bin. The final UV LF is calculated by combining all galaxies with their associated corrected effective volume. Regarding uncertainties, we use the HFF volume uncertainties for the HFF sources, and the updated volume uncertainties for the new sample, respectively. We note that the HFF uncertainties include systematic effects derived from a comparison between four independent models \cite{atek18}. For the three sources outside the HFF coverage, the comparison to another independent model \cite{bergamini23} shows that the uncertainties are negligible for such small amplification factors. We also include Poisson errors and cosmic variance in the final LF results. For the survey volume probed by our program, we estimated cosmic variance to be around $\sigma_{\rm CV} \sim 30\%$ \cite{trenti08}. Overall, In Figure 7, we show that our measurements (orange points) are in good agreement with the faint-end of the photometric UV LF derived from HFF observations\cite{atek18} (gray points).

\begin{figure*}
    \centering
    \includegraphics[width=10cm]{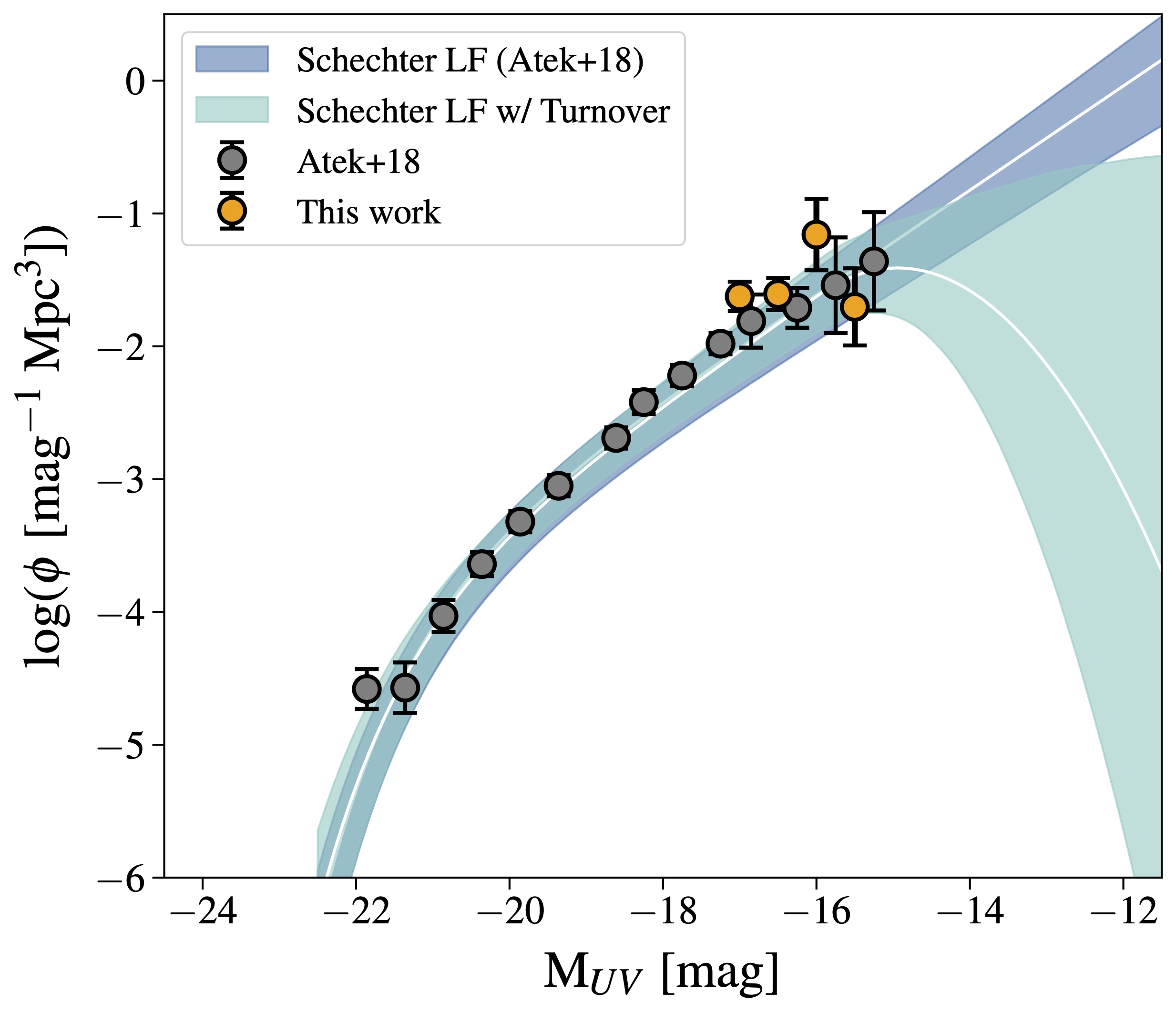}
    \caption{{\bf Spectroscopic constraints on the UV luminosity function}. The UV luminosity function as determined from our spectroscopic sample is represented by orange points. Also shown, the photometric determination from the HFF data \cite{atek18}, together with the best-fit Schechter function (blue curve) and a modified Schechter with a potential turnover (teal curve). The shaded region of each curve represent the 1$-\sigma$ uncertainties }
    \label{fig:enter-label}
\end{figure*}

\noindent In our efforts to determine whether galaxies can reionize the Universe, we proceed to calculate another crucial parameter: the production efficiency of ionizing radiation $\xi_{\mathrm{ion}}$. This quantity is defined as the ratio between the LyC photon production rate in the units of s$^{-1}$, and the observed non-ionizing UV luminosity density $L_{\mathrm{UV}}$ estimated at 1500 \AA\ in 
units of erg s$^{-1}$ Hz$^{-1}$:

\begin{equation}\label{eq:xi}
    \xi_{\rm ion} = \frac{N(H^{0})} {L_{UV}} ~[erg^{-1}~ Hz],
\end{equation}

\noindent  where $N(H^{0})$ can be estimated from the H$\alpha$\ Balmer line\cite{leitherer95} assuming a case B recombination theory\cite{osterbrock89}: 
\begin{equation}\label{eq:nh}
    L(H\alpha)~[erg ~s^{-1}] = 1.36 \times (1-f_{esc})~ 10^{-12} ~N(H^{0}) ~[s^{-1}]  
\end{equation}

\noindent where $ L(H\alpha)$ is in units of erg s$^{-1}$, and $f_{\rm esc}$\ is the escape fraction of Lyman continuum radiation. In this calculation, we assume that $f_{\rm esc}$=0, meaning that all Lyc photons are reprocessed into the Balmer lines. The derived $\xi_{\mathrm{ion}}$\ value can be considered as a lower limit, since higher $f_{\rm esc}$\ will lead to a higher $\xi_{\mathrm{ion}}$. The H$\alpha$\ emission line is not detected in source ID 18924. In this case, we use the H$\beta$\ luminosity and a case B conversion factor. Our measurements are reported in Figure 3, together with literature results \cite{atek22,bouwens16,matthee17,nanayakkara20,matthee23,sun23}. These uncertainties due to potential field-to-field variations can also affect the ionizing properties of galaxies. We incorporated cosmic variance errors ($\sigma_{\rm UV}\sim 30\%$) to the $\xi_{\mathrm{ion}}$\ values for both $M_{\rm UV}$$>-16.5$ mag and $M_{\rm UV}$$<-16.5$ mag (cf. Figure 3).

\noindent Spectroscopic measurements in brighter galaxies have also reported $\xi_{\mathrm{ion}}$\ values higher than canonical values \cite{tang23}. Several {\em JWST}\ studies have measured $\xi_{\mathrm{ion}}$\ in faint galaxies at the epoch of reionization \cite{saxena23, simmonds23}. However, their results are based on emission line fluxes inferred from broadband excess, rather than direct spectroscopic measurements. In particular, {\em JWST}\ medium-band photometric data have been used to infer $\xi_{\mathrm{ion}}$\ for galaxies over the redshift range $3<z<7$ \cite{prieto23}. These measurements offer an opportunity to explore a larger population of galaxies through wide-area imaging. They report ionizing efficiencies in the range log($\xi_{\mathrm{ion}}$/ Hz erg$^{-1}$) =$25.31 - 25.39$, where galaxies with strong Ly$\alpha$ emission tend to have the highest values. These values are smaller than our average measurements for the faintest galaxies. However, the vast majority of their sample is at significantly lower redshifts. Their redshift distribution has two peaks at $z=3$ and $z=5$, and the median redshift of their sample is $z=4.02$, which is well below the epoch of reionization. Among their sample of 370 galaxies, only $\sim 25$ galaxies lie within the epoch of reionization. Therefore, their average $\xi_{\mathrm{ion}}$\ value is not representative of the epoch of reionization. Furthermore, this difference is smaller if we take into account the dynamical range of UV magnitudes explored in their study. Although they find a weak dependency of $\xi_{\mathrm{ion}}$\ with $M_{\rm UV}$, their sample consists of galaxies with UV magnitudes ranging from -23 to -15.5 mag, compared to our subsample of $M_{\rm UV}$$>-16.5$ mag. While imaging-based measurements can be complementary to spectroscopic studies, they have larger uncertainties (in the range $\sigma=$0.43 - 0.64), due to the way emission line fluxes are inferred, and are model-dependent, since the continuum is derived from SED-fitting. 
 
\noindent Finally, we derive constraint on the LyC escape fraction using indirect indicators calibrated in a large sample of nearby LyC emitting galaxies. Since the LyC emission is impossible to measure at the epoch of reionization, large efforts have been devoted in the last two decades to determine the escape fraction in $z<4$ galaxies, and more importantly establish indirect methods, which can be transferable to reionization sources. This was precisely the motivation of the recent Low-redshift Lyman continuum survey (LzLCS)\cite{flury22a}. Among the different physical properties of LyC leakers, the observed UV continuum slope $\beta$ has been identified as a promising proxy of $f_{\rm esc}$ \cite{chisholm22}. Here, we use the UV slopes derived from the best-fit models of {\tt Bagpipes} and the LzLCS relation to infer $f_{\rm esc}$. The derived values for the present sample range from 4.5\% to 15.6\%. Despite large uncertainties (around 50\%), only two sources have $f_{\rm esc}$\ values that can reach below 4\% at 1-sigma.  

\noindent Now with all three properties in hand, we can assess the contribution of faint galaxies to cosmic reionization. We compute the ionizing photon emissivity of galaxies, which is the product of the total UV luminosity density $\rho_{\mathrm{UV}}$, derived from integrating the UV LF, and the production efficiency $\xi_{\mathrm{ion}}$. The result will depend on the faint integration limit, which is set the faintest bin of the spectroscopic LF at $M_{\rm UV}$=-15 mag.  By multiplying this quantity by the escape fraction $f_{\rm esc}$, we obtain the total ionizing photon rate density that is available to ionize the IGM. Down to $M_{\rm UV}$=-15 mag, modest $f_{\rm esc}$ values around 5\% are sufficient to maintain reionization.

\noindent We measure the gas-phase oxygen abundance using the strong optical lines diagnostic. Specifically, we use the R3=log([O{\sc iii}]$\lambda$5007 /H$\beta$) and adopt the most recent empirical calibrations at high-redshift\cite{nakajima22,nakajima23,sanders21}. All of our sources show detections of [O{\sc iii}]$\lambda$5007 and H$\beta$. Furthermore, we also separate our sample into two bins according to the EW(H$\beta$) at EW(H$\beta$)=100\AA\ to account for ionization parameter variations\cite{nakajima22}. Also, for a given value of R3, the calibration defines two metallicity solutions. Although these sources have likely low metallicities, we use the O32=log([O{\sc iii}]$\lambda\lambda$4949,5007/[O{\sc ii}]$\lambda\lambda$3727,3729) ratio to distinguish between the two branches. For most of the sources, the [O{\sc ii}] is not detected, which provides a lower-limit on O32, which is found to vary in the range O32=[0.7-1.4]. Using the O32 metallicity indicator, the resulting values are all compatible with the low-metallicity branch solution. The metallicity measurements are reported in Table 2. 

\noindent In addition to the physical properties that may ease the escape of ionizing photons from these galaxies, metallicity can also inform us on the ionizing properties of this population. We measured the gas-phase metallicity using the R3 = [O{\sc iii}]/H$\beta$\ line ratio based on the most recent calibrations \cite{sanders23}. We find extremely low metallicities, ranging from 12+log(O/H)= 6.70 to 7.46, which corresponds to 1\% to 6\% of the solar metallicity. Such low metallicities are often suggestive of strong ionizing radiation from massive stars\cite{stanway19}. At the same time, these distant low-mass galaxies, considered as the building-blocks of present-day galaxies, are expected to be metal-poor, owing to their supposedly pristine gas conditions. We note that our estimate relies on the calibration of strong lines diagnostics, which are prone to significant uncertainties at $z>6$. On average, these estimates lie with 1 to 2$\sigma$ intervals from each other.

\subsection*{Size Measurements}

To further characterize these sources, we measure their sizes by fitting their morphology in the NIRCam F150W filter with a S\'{e}rsic profile\cite{sersic63}. Measurements of the half-light radii for all eight galaxies are performed using the \texttt{pysersic} \cite{Pasha2023} package. For each source, we assume a single Sersic profile and mask all nearby sources in the photometric catalog\cite{weaver23}. The priors for the half-light radius, Sersic index,  axis ratio and position angle are all uniform and varied from $0.5 - 10$ pixels, $0.65 - 4$, $0.1 - 1$ and $0 - 2\pi$ respectively. Priors central position and flux are represented as Gaussian distributions with the location and width based on the photometric catalog\cite{weaver23}. A flat sky background is fit simultaneously. The Posterior distribution is explored using the No-U Turn (NUTS) \cite{Hoffman2014} sampler implemented in \texttt{numpyro} \cite{Phan2019} with 2 chains for 1,000 warm-up and sampling steps each. 

\noindent Several of our objects are significantly distorted by the gravitational lensing which means that we need to take the shear into account when deriving the half-light radius. In order to do that, we use our lensing model to derive the tangential and radial magnifications, defined as $\mu=\mu_{\mathrm{t}}\mu_{\mathrm{r}}$, from the deflection field at each source's position and redshift. Since our objects are sheared along the tangential direction, we use the tangential component of the magnification to correct the half-light radii.

\noindent The derived effective radii, corrected for magnification and taking the shear into account, vary between $r_{\rm eff}$= 30 to 300 pc. Overall, these constraints show that these sources are small, in broad agreement with an extrapolation to lower masses of the size-mass relation derived at similar redshifts\cite{holwerda15}, albeit with significant scatter. Such small sizes and high sSFRs also supports the scenario of stochastic star formation histories in these systems or dust ejection \cite{ferrara22}, owing to their small dynamical time and a low gravitational potential.

\bmhead{Data Availability}
The NIRCam and {\em HST}\ imaging data are available on the UNCOVER webpage: \url{https://jwst-uncover.github.io/}. 
The NIRSpec spectroscopic data are publicly available through the \texttt{Mikulski Archive for Space Telescopes} (\texttt{MAST}; \url{https://archive.stsci.edu/}), under program ID 2561. The UNCOVER Lensing products are available at \url{https://jwst-uncover.github.io/DR1.html##LensingMaps}.

\bmhead{Code Availability}
Astropy \cite{astropy13, astropy18}, 
Bagpipes \cite{Carnall2018,Carnall2019:VANDELS}, 
BEAGLE \cite{chevallard16},  
EAzY \cite{Brammer2008}, 
Matplotlib \cite{plt07}, 
msaexp v0.6.10 \cite{brammer22msa}, 
NumPy \cite{numpy20}, 
NUTS \cite{nuts11,nuts19}, 
PyMultinest \cite{Buchner:2014, Feroz:2019}, 
pysersic \cite{Pasha2023}, 
SciPy \cite{scipy20},
GrizLi \url{https://github.com/gbrammer/grizli}

\bmhead{Acknowledgments}
H.A. and IC acknowledge support from CNES, focused on the JWST mission, and the Programme National Cosmology and Galaxies (PNCG) of CNRS/INSU with INP and IN2P3, co-funded by CEA and CNES. H.A thanks the Cosmic Dawn Center (DAWN) for their support. DAWN is funded by the Danish National Research Foundation under grant No. 140. IL acknowledges support by the Australian Research Council through Future Fellowship FT220100798. P.D. acknowledges support from the NWO grant~016.VIDI.189.162 (``ODIN") and from the European Commission's and University of Groningen's CO-FUND Rosalind Franklin program. A.Z. acknowledges support by Grant No. 2020750 from the United States-Israel Binational Science Foundation (BSF) and Grant No. 2109066 from the United States National Science Foundation (NSF), and by the Ministry of Science \& Technology, Israel. 
The work of C.C.W. is supported by NOIRLab, which is managed by the Association of Universities for Research in Astronomy (AURA) under a cooperative agreement with the National Science Foundation.

\bmhead{Author Contributions}
H.A. led the analysis and article writing. 
L.J.F. and A.Z. constructed the lens model and extracted lensing related quantities. S.F produced figures. I.L. and R.B. are the PIs of the UNCOVER program. R.B. and I.L. designed the observations and reduced the spectra. J.W. and B.W. produced the catalogs used for target selection. P.D. provided simulations to interpret the observational results obtained. V.K. produced line measurements.  I.C. estimated survey volumes. D.J.S. ran SED fitting analysis. T.B.M. measured the galaxy sizes. All authors contributed to the manuscript and aided the analysis and interpretation.

\bmhead{Author Information}
Correspondence and requests for materials should be addressed to \url{hakim.atek@iap.fr}. Reprints and permissions information is available at \url{www.nature.com/reprints}.
The authors declare no competing interests.

\end{document}